      [1999/12/01 v1.4c Il Nuovo Cimento]
\documentclass{cimento}

\usepackage{graphicx} % got figures? uncomment this 

\title{The search for optical emission on and before the GRB trigger
with the WIDGET telescope}
\shorttitle{The search for optical emission on and before the GRB trigger}
\author{
T.~Tamagawa\from{ins:RIKEN}\ETC,
F.~Usui\from{ins:ISAS},
Y.~Urata\from{ins:RIKEN}\from{ins:TITECH},
K.~Abe\from{ins:SAITAMA},
K.~Onda\from{ins:RIKEN}\from{ins:TUS},
M.~Tashiro\from{ins:SAITAMA},
Y.~Terada\from{ins:RIKEN},
H.~Fujiwara\from{ins:UT},
N.~Miura\from{ins:UT},
S.~Hirose\from{ins:UT},
N.~Kawai\from{ins:TITECH}\from{ins:RIKEN},
A.~Yoshida\from{ins:AGU}\from{ins:RIKEN},
M.~Mori\from{ins:ICRR},
K.~Makishima\from{ins:UT}\from{ins:RIKEN}
}

\instlist{
\inst{ins:RIKEN} RIKEN, Institute for Physical and Chemical
Reserch, Wako, Saitama 351-0198, Japan
\inst{ins:ISAS} Japan Aerospace Exploration Agency, Institute of Space
and Astronautical Science, Sagamihara, Kanagawa 229-8510, Japan.
\inst{ins:TITECH} Tokyo Institute of Technology, Ookayama, Meguro, Tokyo
152-8550, Japan.
\inst{ins:TUS} Tokyo University of Science, Kagurazaka, Shinjyuku, Tokyo
162-8601, Tokyo.
\inst{ins:SAITAMA} Saitama University, Sakura-ku, Saitama 338-8570,
Japan.
\inst{ins:AGU} Aoyama Gakuin University, Sagamihara, Kanagawa 229-8558,
Japan.
\inst{ins:ICRR} Institute for Cosmic Ray Research, University of Tokyo,
Kashiwa, Chiba 277-8582, Japan.
\inst{ins:UT} University of Tokyo, Bunkyo-ku, Tokyo 113-0033, Japan.
}
\PACSes{
\PACSit{95.55.Cs}{Ground-based ultraviolet, optical and infrared telescopes}
\PACSit{95.55.Qf}{Photometric, polarimetric, and spectroscopic instrumentation}
\PACSit{98.70.Rz}{gamma-ray sources; gamma-ray bursts}
}
\begin{document}

\maketitle

\begin{abstract}
{\it WIDGET} is a robotic telescope for monitoring the {\it HETE-2}
field-of-view to detect Gamma-ray Burst optical flashes or possible
optical precursors. The system has $62^\circ\times62^\circ$ wide
field-of-view which covers about 80\% of {\it HETE-2} one with a
2k$\times$2k {\it Apogee} U10 CCD camera and a {\it Canon} EF 24mm
f/1.4 wide-angle lens without a bandpass filter. {\it WIDGET} has been
in operation since June 2004 at Akeno observing site where is about
200 km apart from Tokyo. Typical limiting magnitude with S/N=3 at the
site is V=10$^{mag}$ for 5 seconds exposure and V=11$^{mag}$ for 30
seconds exposure. We had already six coincident observations with {\it
HETE-2} position alerts. It was, however, cloudy for all cases due to
rainy season in Japan. Expected number of coincident observations
under clear sky is about 5 events per year. We will extend the system
in early 2005 for {\it Swift} era to monitor optical transients in
wider field-of-view, multi-color or polarization modes.
\end{abstract}

\section{Introduction}

Detection of optical emission on and before Gamma-Ray Burst (GRB)
triggers provide us knowledge of gamma-ray emission mechanism at early
stage of an ejecta expansion from a massive star explosion. The optical
flash can be produced by a reverse shock in the ejecta \cite{sari_1999},
or produced from loaded electron-positron plasma by prompt gamma-ray
emission\cite{beloborodov_2002}\cite{paczynski_2001}\cite{fan_and_wei_2004}.
Although there is much effort to search such prompt optical emission
from GRB, only one flash has been detected as far; during the GRB 990123
was active, very bright optical flash was observed by ROTSE-I telescope
at 8.9$^{th}$ magnitude about 50 seconds after the burst trigger
\cite{akerlof_1999}.

Recently several groups are developing and operating wide-field
telescopes for hunting the optical flashes; {\it e.g.}  RAPTOR
\cite{RAPTOR}, BOOTES \cite{BOOTES}, Ashra \cite{Ashra}, WFOC
\cite{WFOC_InThisPaper}, $\pi$ of the Sky \cite{PiOfTheSky} and so
on. We also started construction of a wide-field telescope named {\it
WIDGET} (Wide-field telescope for GRB early timing) in late 2003.

\section{The {\it WIDGET} System}

The {\it WIDGET} telescope consists of an {\it Apogee} Alta U10 CCD
camera and a 24 mm f/1.4 {\it Canon} EF wide-angle lens. The U10 camera
uses a 2048 $\times$ 2048 format {\it Atmel} THX7899 CCD chip with
14-micron pixels. The chip is front-illuminated type with quantum
efficiency of 38\% at 720nm.  The camera provides a
$62^\circ\times62^\circ$ field of view (FOV). Electronics onboard the
CCD camera have been optimized to provide fast readout of the entire
array in 5 seconds with a 16 bit AD converter through an USB2.0 serial
line. The standard exposure in our operation is 5 seconds, so that we
take sky images every 10 seconds. The optics and the CCD camera are
attached to a {\it Takahashi} NJP Temma-2 polar mount. The mount is
capable of slewing at a speed of about 90 arcmin/sec. It is controlled
by an PC running Linux OS via a serial line.

\begin{figure}
\includegraphics[width=0.5\textwidth]{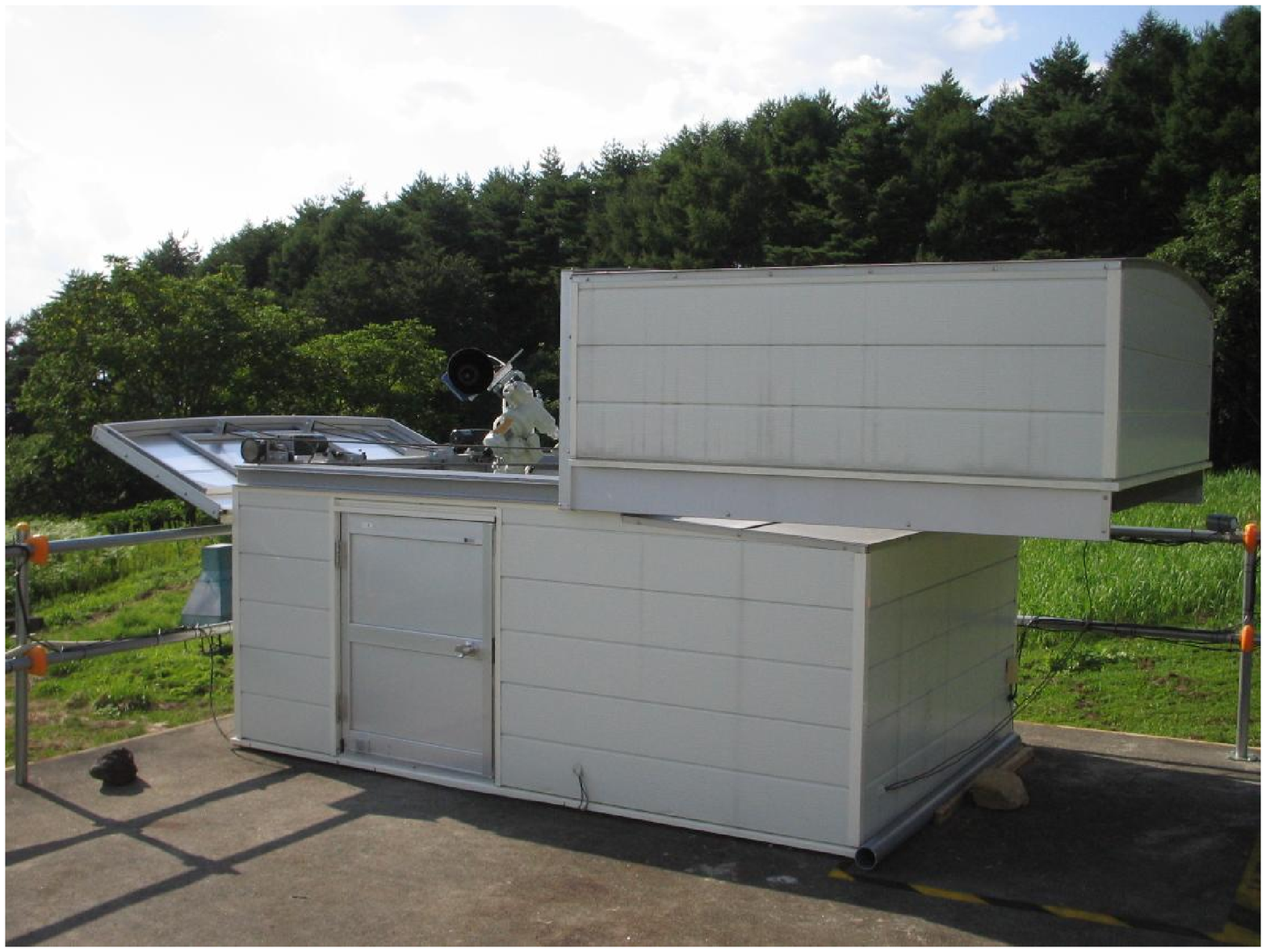}
\includegraphics[width=0.5\textwidth]{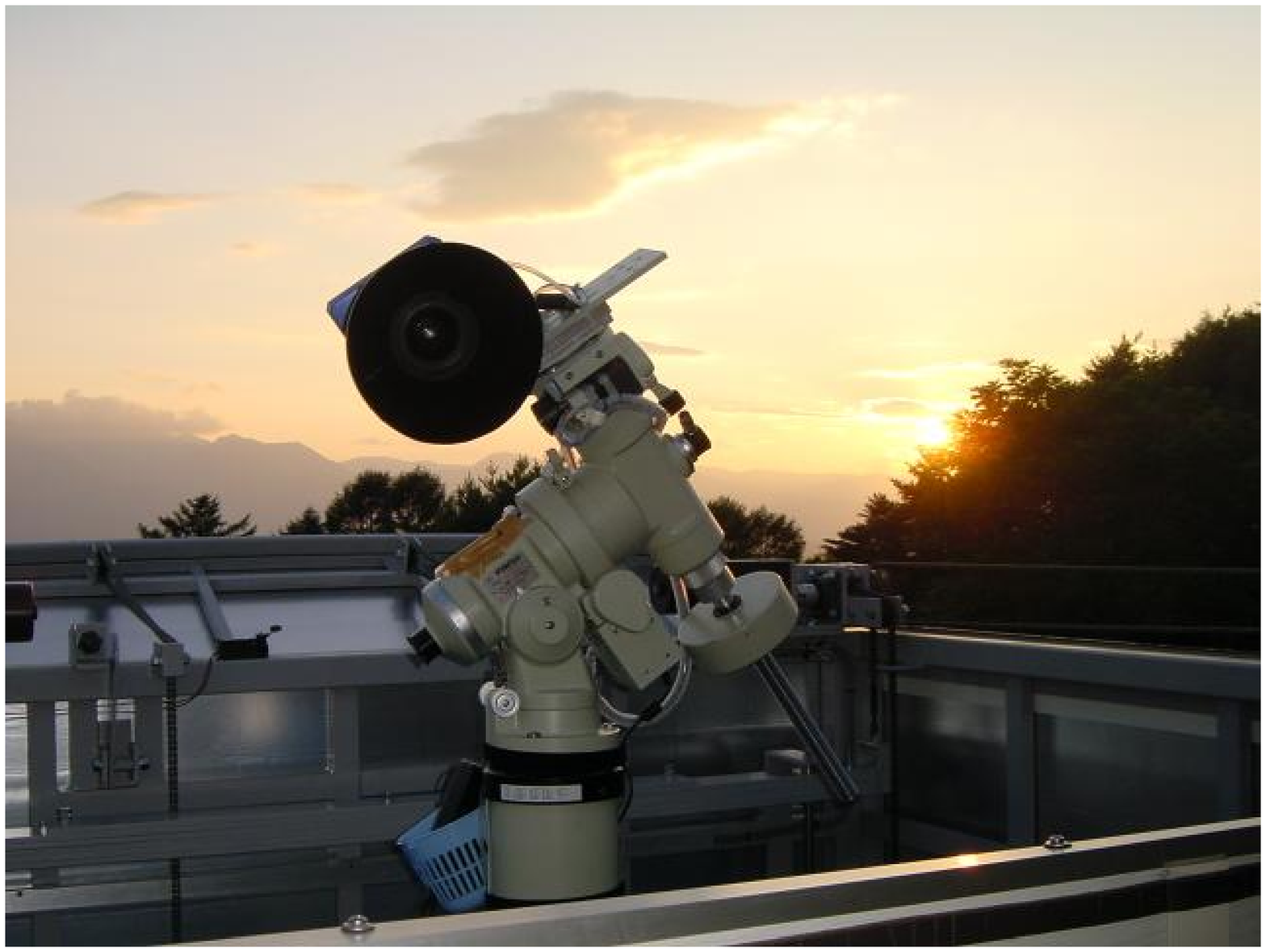} \caption{(Left) The {\it
WIDGET} enclosure hut with a sliding roof. (Right) The CCD camera and
the wide-angle lens attached on the polar mount housed inside the hut.}
\label{fig:widget}
\end{figure}

The telescope, related computers and electronics are housed inside the
hut whose roof is sliding to one direction. It is custom build by {\it
Human Comm Co. Ltd.} The height and width are 2 meters, and the length
is 3 meters (see figure \ref{fig:widget}). Control of the roof is
provided by relay boards attached on a side-wall of the hut. The relay
boards accept commands via a socket connection from a control PC
running Linux. The roof is opened and closed according to a preplanned
schedule. In case of rain the roof is closed by a circuit connecting
to weather stations located outside the hut. When a power outage
occurs, the roof is automatically closed by weight of some steel bars
without any electrical power. The timing of the whole system is
provided by a network timing protocol (NTP) server and a radio clock
with an average error less than 1 second.

\begin{figure}
\includegraphics[width=0.61\textwidth]{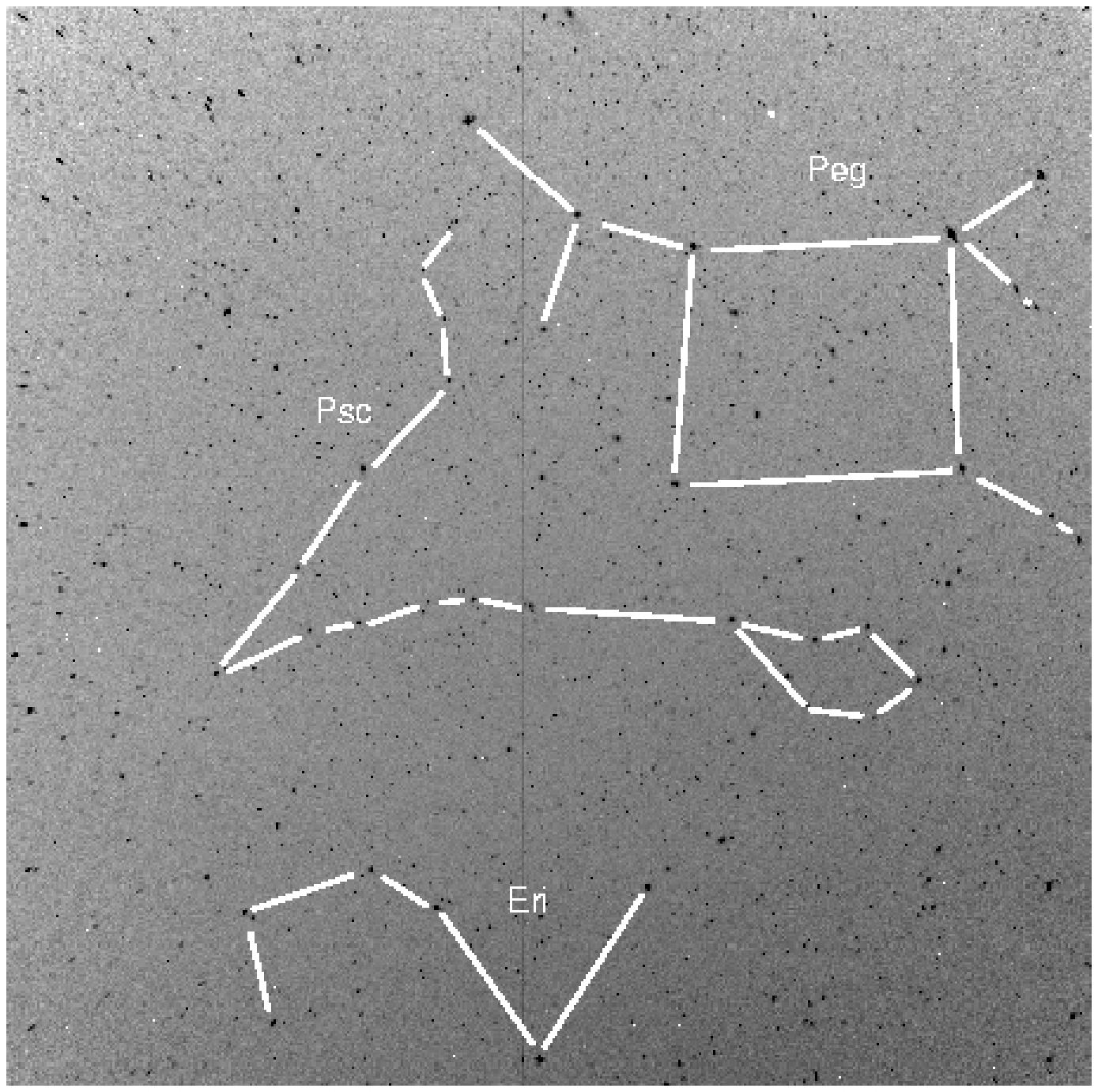}
\includegraphics[width=0.39\textwidth]{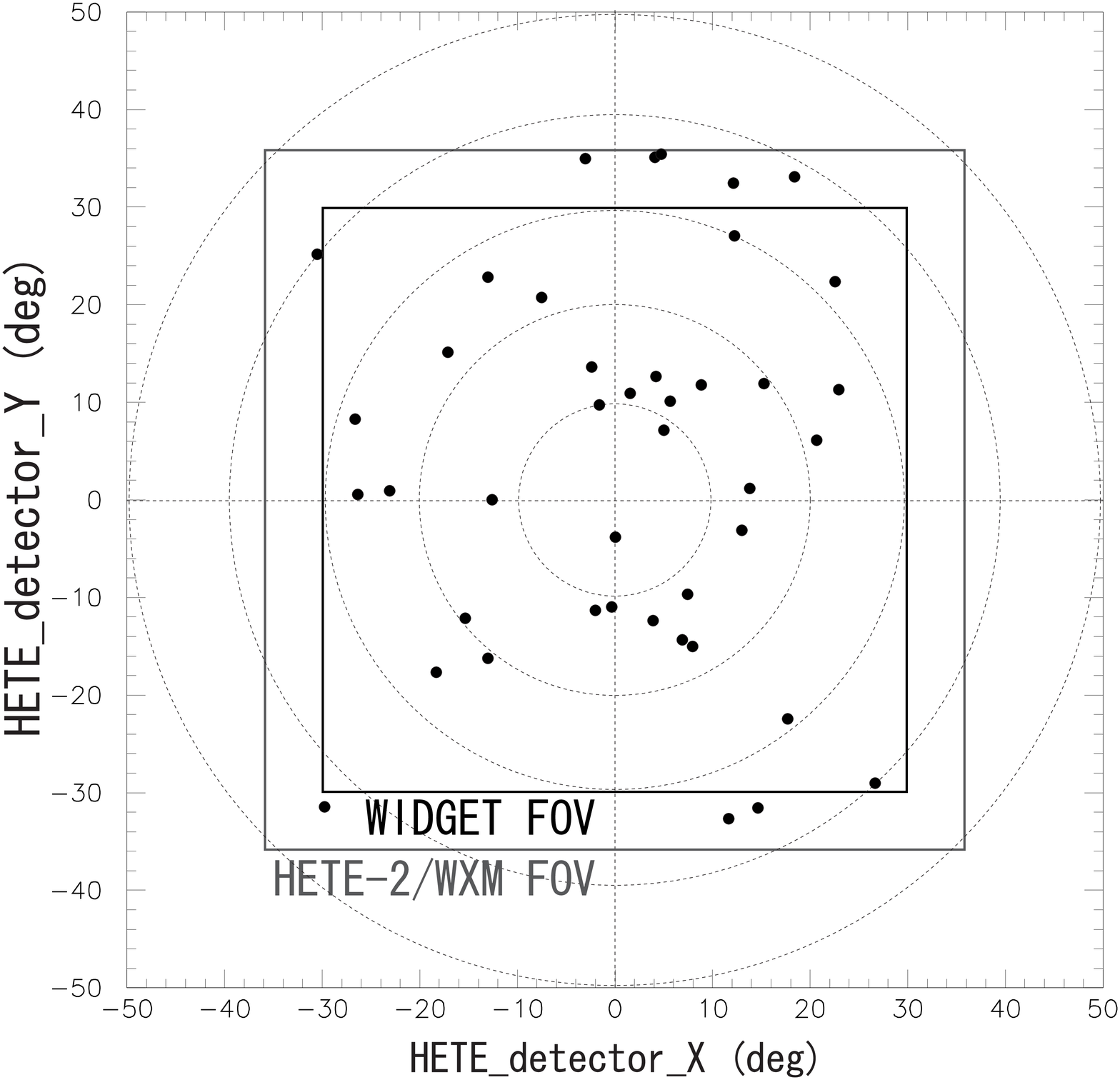}
\caption{(Left) A sky image taken by WIDGET measuring 62$^\circ$ on a
side. (Right) The comparison of the WIDGET FOV (inner rectangle) and the
HETE-2 FOV (outer rectangle) in the HETE-2 detector coordinate. Dots in
the image denote the GRB positions reported by HETE-2 in the last three
years \cite{Sakamoto_2004}.} \label{fig:sky_image}
\end{figure}

The {\it WIDGET} telescope is placed at the Akeno campus which belongs
to the Institute for Cosmic Ray Research, University of Tokyo. The site
was used for the air shower experiment, Akeno Giant Air Shower Array
(AGASA) \cite{AGASA}. Since we design the system for fully automatic
operation, almost all of the devices are robust against software errors,
hardware failures, or network troubles. We started observations at the
site in June 2004, and established the fully automatic operation in
September 2004. The station is currently unmanned except for regular
maintenance.

\section{System Performance}

Astrometric and photometric calibrations are based on the standard stars
in the Tycho catalog \cite{TYCHO}. For calibration and monitoring of
data quality, we uses more than 10 well calibrated standard stars. The
limiting magnitude is about $V=10^{mag}$ for 5 seconds exposure and
$V=11^{mag}$ for 30 seconds exposure with S/N=3 at the Akeno observing
site. The limiting magnitude is somewhat reduced when the moon is in the
FOV. Stability of the magnitude of a star measured by the photometry is
about $\Delta V=0.2^{mag}$ in a night.

Because of the optics distortion, we can not simply fit a celestial
coordinate to the whole image. To reduce the distortion effect, we
restrict the field for astrometry in the $10^\circ\times10^\circ$ region
which enclose the GRB position reported by a satellite or ground based
observations. As the result, we set the astrometric precision about 1.5
arcseconds even around the edge of the {\it WIDGET} FOV. Since the
precision is well below a size of one pixel, we can identify the
position of GRBs on the chip without any ambiguity.

\section{Current Status and Future Plans}

Since June 2004 where we started scientific operation, we have carried
out six coincident observations with {\it HETE-2} position alerts
listed on the table \ref{tab:grb_detection}. Unfortunately, we missed
all of the chances due to 2004's bad weather condistions in Japan. We
continue the automatic observation when it is fine. About five
coincident observations are estimated per year under clear sky.

We are extending the system in early 2005 for the {\it Swift} era. One
of the extensions is to add three more telescopes on the
mount. Totally four telescopes cover the larger sky area of about
$100^\circ\times100^\circ$ which is fair part of the {\it Swift}
FOV. The other option is to add some bandpass filters for measuring
color variation of the GRB optical flashes. Although the present
system is not capable of mounting the filters, we can install them by
remodeling of the camera attachment. We expects that the color
variation will probably tell us physical conditions or origin of the
optical flashes. The other extension is to add three polarized filters
with polarization angles of 0$^\circ$, 60$^\circ$, and 120$^\circ$ for
measuring polarization of the optical transients. High liner
polarization in the optical flashes may be detected in some
GRBs\cite{fan_and_wei_2004}.

\begin{table}
 \caption{WIDGET GRB observations coincident with the HETE-2 position
 alerts.}  \label{tab:grb_detection}
 \begin{tabular}{ccccc}
  \hline
  GRB & Time (UT) & $\theta_{GRB}$ $^\dagger$ & WIDGET status & Results \\
  \hline
  040810  & 14:15:35 & 28$^\circ$ & running & cloudy \\
  040825B & 16:21:37 & 9$^\circ$  & running & cloudy \\
  040912  & 14:12:16 & 7$^\circ$  & running & cloudy \\
  040924  & 11:52:11 & 34$^\circ$ & running & cloudy \\
  041006  & 12:18:08 & 6$^\circ$  & running & cloudy \\
  041211  & 11:31:51 & 13$^\circ$ & running & cloudy \\
  \hline
  \multicolumn{5}{l}{$\dagger$ Angle between GRB position and the centre
  of the WIDGET FOV}\\
 \end{tabular}
\end{table}

\acknowledgments 

This project is supported by RIKEN Director's Fund in FY2003 and FY2004.

\end{document}